\documentclass[prd,epsf,aps,twocolumn,letterpaper,preprintnumbers]{revtex4}
\usepackage{amssymb}
\usepackage{amscd}
\usepackage{amsfonts}
\usepackage{epsf}
\usepackage{graphicx}
\begin{document}
\title{Inflation and Oscillations of Universe in 4D Dilatonic Gravity}
\author{ P.~Fiziev,
\footnote{
    Electronic address: {\tt fiziev@phys.uni-sofia.bg}}
        D.~Georgieva%
\footnote{
    Electronic address: {\tt daniag@phys.uni-sofia.bg}
}
}
\address{
        Department of Theoretical Physics,
    University of Sofia, 5 James Bourchier Blvd.,
    1164 Sofia, Bulgaria.}

\begin{abstract}
We investigate the inflation of Universe in a model of four
dimensional dilatonic gravity with a massive dilaton field $\Phi$.
The dilaton plays simultaneously the roles of an inflation field
and a quintessence field. It yields a sequential {\em
hyper}-inflation with a graceful exit to asymptotic de Sitter
space-time, which is an attractor, and is approached as
$\exp(-\sqrt{3\Lambda^{obs}}\,ct/2)$. The time duration of the
inflation is reciprocal to the the mass of the dilaton: $\Delta
t_{infl}\sim m_{{}_\Phi}^{-1}$. The typical number of e-folds in
the simplest model of this type is shown to be realistic without
fine tuning.

\noindent{PACS number(s): 04.50.+h, 04.40.Nr, 04.62.+v}
\end{abstract}
%%%%%%%%%%%%%%%%%%%%%%%%%%%%%%%%%%%%%%%%%%%%%%%%%%%%%%%%%%%%%%%%%%%
\maketitle
%
%\draft
\sloppy
%\scrollmode
%%%%%%%%%%%%%%%%%%%
\newcommand{\lfrac}[2]{{#1}/{#2}}
\newcommand{\sfrac}[2]{{\small \hbox{${\frac {#1} {#2}}$}}}
\newcommand{\ben}{\begin{eqnarray}}
\newcommand{\een}{\end{eqnarray}}
\newcommand{\la}{\label}
\newcommand{\BBox}{{\square}}
\newcommand{\Si}{{\rm Si}\,}
%
%%%%%%%%%%%%%%%%%%%%%%%%%%%%%%%%%%%%%%%%%%%%%%%%%%%%%%%%%%%%%%%%%%%%%%%%%
\section{Introduction}

We consider a specific 4D-dilatonic model of gravity (4D-DG) with
a gravi-dilaton action \cite{F00,F02}: \ben {\cal A}_{g,\Phi}=
-{\frac c {2\kappa}}\int d^4 x\sqrt{|{\bf g}|}
 \bigl(\Phi R + 2 \Lambda^{obs}\, U(\Phi) \bigr),
\la{4A} \een where $\kappa=8\pi G_N/c^2$ is the Einstein constant and
 $\Lambda^{obs}$ is the observed value of the cosmological
constant. In our model we have only one unknown function -- the
{\em cosmological potential} $U(\Phi)$ -- which has to be chosen
to comply with {\em all} gravitational experiments and
observations in laboratory experiments, in star systems, in astrophysics and
cosmology. In addition it has to solve the {\em inverse
cosmological problem}, namely, to determine $U(\Phi)$ that
reproduces the  time evolution of the scale parameter $a(t)$ in accordance with the
Robertson-Walker (RW) model of Universe \cite{F00,F02}). Further
on we use {\em cosmological units}, in which the cosmological
constant $\Lambda^{obs}$ equals one \cite{F00,F02}.

The dilaton field $\Phi$ is a quantity, reciprocal to the
dimensionless gravitational factor $g$:\,\, $\Phi=1/g$. As usual
in our scalar-tensor theory of gravity of Brans-Dicke type,
instead of the Newton gravitational constant $G_{N}$, we define
the variable gravitational factor $G_{eff}(x)=G_{N}\,g(x)$ as a
function of space-time coordinates $x=\{ct,{\bf x}\}$.

There exist different reasons for considering this model. For
example, it appears in the five-dimensional Kaluza-Klein model, or, it
can be produced in the low energy limit of string theory, by ignoring
(yet unobserved) possible higher dimensions. (See for details and
for references \cite{F02}.) The most physically relevant motivation for the
use of a model with action (\ref{4A}) at present can be find in
the attempts to develop a quantum version of Einstein gravity in
four dimensional space-time \cite{QNLG}.

Recently, a very promising discovery of the existence of a
consistent quantum gravity in four dimensional space-time was made
by Reuter \cite{Reuter}. It confirms an early prediction by
Weinberg that Einstein gravity in four dimensions may be
"asymptotically safe", because of the existence of a fixed point
of the renormalization-group equations, similar to the one in
$2+\varepsilon$ dimensions \cite{Weinberg79}. Indeed, by making
use of suitable Einstein-Hibert truncation, i.e., using action
(\ref{4A}) with $\Phi=1$ and $U(\Phi)=1$  it has been proven that
the four-dimensional possesses an  ultra-violet fixed point and is
asymptotically free. Then, as a result of the renormalization, one
obtains a running gravitational constant $G$ and a cosmological
constant $\Lambda$. In the case of quantum electrodynamics we have
running electric charges, due to the cloud of virtual photons. In
quantum gravity, a test body would be surrounded by a cloud of
virtual gravitons. In contrast to electrodynamics though, the
gravity is universal attractive and the effective mass of the
body, as seen by a distant observer, is larger than it would be in
absence of quantum effects. Hence, in quantum gravity we have {\em
an anti-screening} effect, which entails the Newton constant and
the cosmological constant becoming scale dependent quantities. As
a result of the general relativistic invariance they become
functions of space-time points.

We prefer to describe the above phenomena using the dilaton field
$\Phi$ in action (\ref{4A}). This field encapsulates the
space-time dependence of the corresponding parameters. The use of
a cosmological potential $U(\Phi)$ of general form is known to
be equivalent to the nonlinear gravity (NLG), described by a
nonlinear (with respect to the space-time scalar curvature $R$)
Hilbert-Einstein Lagrangian $L(R)$ in the action. (See \cite{F02, NLG}
for more details and for references.) At present, the form of this
Lagrangian is complete unknown. The renorm-group analysis of the
nonlinear terms in Reuter approach is still an open problem. It is
clear that, in general, all coefficients in the different terms of the
Taylor (or Laurent) series expansion of the function $L(R)$
will become running constants, as a result of quantum effects.

As shown in \cite{F00,F02}, the use of the cosmological potential
$U(\Phi)$, as an alternative description of these nonlinear terms,
is extremely useful. Indeed, one is able to formulate simple and
clear physical requirements for this potential. For example, to
have an unique physical {\em de Sitter} (classical) {\em vacuum}
(dSV) in the theory, the function $U(\Phi)$ must have an unique
positive minimum for positive values of the dilaton field $\Phi$. To
avoid the non-physical negative values of the dilation field $\Phi$,
the potential must increase to infinity for $\Phi \to +0$, etc.
(See \cite{F02}.)

Here we use the simplest one-parameter potential: \ben U(\Phi) =
\Phi^2+W(\Phi) = \Phi^2+{3\over {16}}p^{-2}_{{}_\Phi}
\left(\Phi-{1\over\Phi}\right)^2\, \la{Vsimplest} \een with such
properties. The parameter $p_{{}_\Phi}$ is the dimensionless
Compton length of the dilaton in cosmological units, defined by
the equation $p^2_{{}_\Phi}=\Lambda^{obs}\, l_{{}_\Phi}^2$, where
$l_{{}_\Phi}={\hbar\over{c m_{{}_\Phi}}}$ is the usual Compton
length of the dilaton, and $m_{{}_\Phi}$ is the mass of the
dilaton. To comply with all known facts in gravitational physics,
astrophysics and cosmology the dilaton mass must obey the
constraint equation $m_{{}_\Phi}> 10^{-3}\rm eV$. This yields an
extremely small value of the dimensionless parameter $p_{{}_\Phi}<
10^{-30}$ \cite{F00,F02}.

The cosmological potential (\ref{Vsimplest}) corresponds to a
non-polynomial Lagrangian for the equivalent NLG:
$L(R;p_{{}_\Phi})\sim -f(R;p_{{}_\Phi})$. It is described in
parametric form by the equations: $R={3\over{4}}p^{-2}_{{}_\Phi}
\left({1/\Phi^3}-\Phi\right)-4\Phi$ and $f={3\over
{8}}p^{-2}_{{}_\Phi} \left({3/\Phi^2}-\Phi^2-2\right)-2\Phi^2$,
$\Phi \in (0,\infty)$. One can see that the {\em physical} branch
of the function $f(R;p_{{}_\Phi})$ has a Taylor series expansion
$$f(R;p_{{}_\Phi})=R+2+1/3\,(R+4)(R +8)\,p_{{}_\Phi}^2+{\cal
O}(p_{{}_\Phi}^4)$$ and goes to the linear Einstein-Hilbert one
$f(R;p_{{}_\Phi})\to R+2$ (in cosmological units) in the limit
$p_{{}_\Phi}\to 0$, i.e., when $m_{{}_\Phi}\to\infty$. Hence, in
this sense, our 4D-DG has the GR as a limiting case.

\begin{figure}[htbp]
\vspace{6.5truecm} \includegraphics{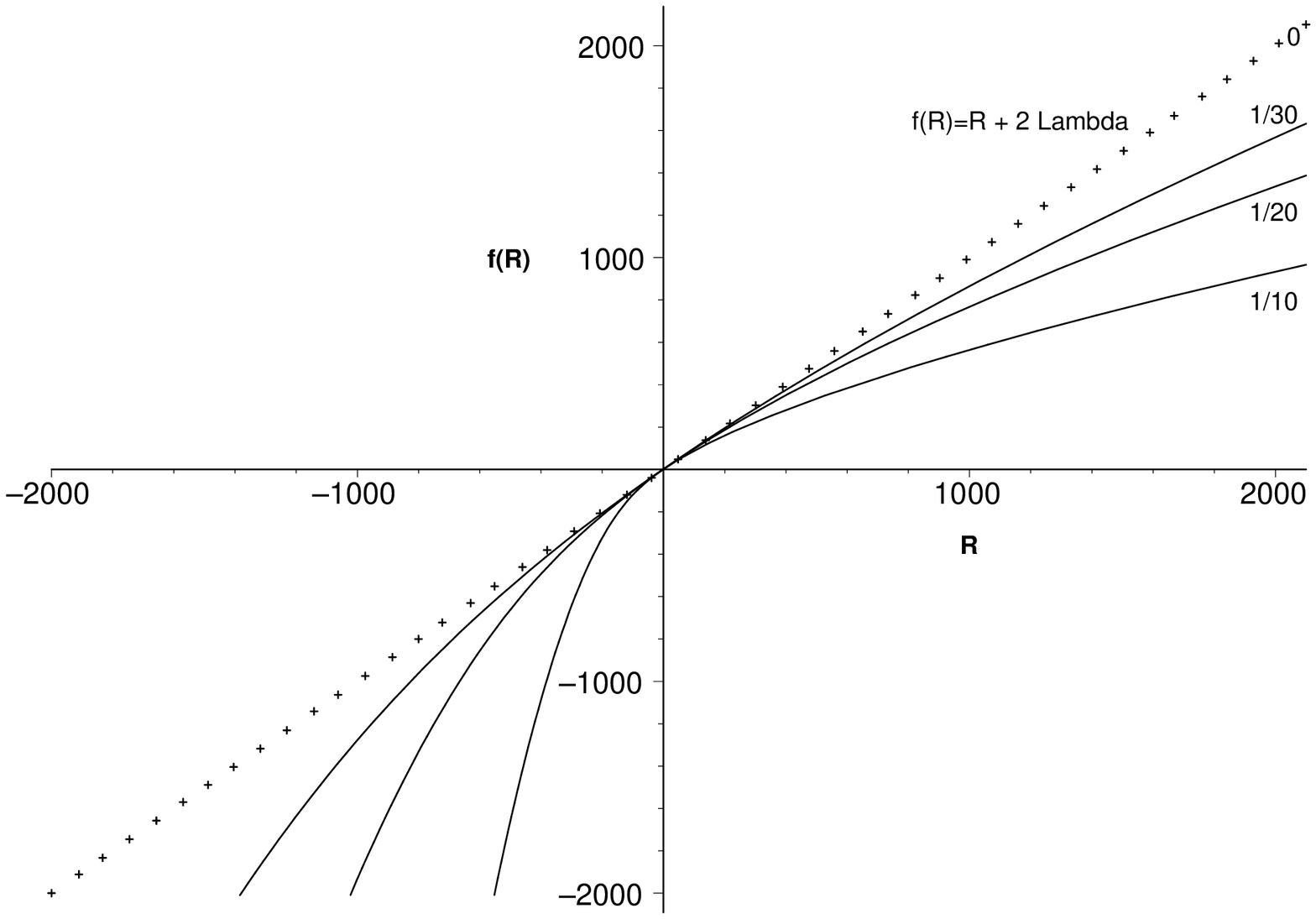}
\caption{\hskip 0.2truecm The form of the function
$f(R;p_{{}_\Phi})$ in the equivalent nonlinear gravity for
different values of the parameter $p_{{}_\Phi}= 1/10, 1/20,
1/30,...,0$.
    \hskip 1truecm}
    \label{Fig0}
\end{figure}

As seen in Fig. \ref{Fig0}, for values of the scalar curvature
$|R|\ll 1$ our model behaves like GR with standard Newton constant
$G_N$. For $R \ll -1$ it behaves like GR with smaller effective
gravitational constant, and for $R\gg 1$ -- like GR with bigger
effective gravitational constant: $$G_{eff}|_{{}_{R \ll\!-1}}<
G_{eff}|_{{}_{|R|\ll\!1}}\approx G_N < G_{eff}|_{{}_{R\gg\!1}}.$$

In the general case, one must consider a 4D-DG model with a matter
sector and a gravi-dilaton sector. The investigation of the effect
of the presence of matter in Reuter approach to quantum gravity was
started in \cite{Percacci}. Some general properties of 4D-DG model
with matter were derived in \cite{F02}. In the present article we
shell ignore the influence of matter for simplicity. Its presence
yields a new important phenomena, which we intend to describe
in a separate article.

\section{The Basic Equations for Robertson-Walker (RW)
         Universe in the 4D-Dilatonic Gravity}

Consider RW adiabatic homogeneous isotropic Universe with
$$ds^2_{RW}= dt^2 - a^2(t) dl^2_k,$$
 where $dl^2_k={\frac {dl^2}{1-kl^2}+l^2(d\theta^2+
\sin^2\theta)d\varphi^2}$ (for $k=-1, 0, 1$). In our cosmological units
the time $t$, and the scale parameter of Universe $a(t)$ are dimensionless.

For a free of matter Universe we obtain the following
basic dynamical equations, governing
the time evolution of the 4D-DG-RW model:
\ben
{\frac {\ddot a} a}+
{\frac {\dot a^2} {a^2}}+ {\frac k {a^2}}=
{\frac 1 3}\, U_{{,_\Phi}}(\Phi),\nonumber\\
{\frac {\dot a} a} \dot\Phi+ \Phi \left({\frac {\dot a^2} {a^2}}
+{\frac k {a^2}}\right)= {\frac1 3}U(\Phi). \la{DERWU} \een The
dynamical  equation for the dilaton $\Phi$ is
\ben
\ddot\Phi+3{\frac{\dot a} a}\dot\Phi+V_{,_\Phi}(\Phi)=0
\la{EqPhi}
\een
and
\ben V(\Phi)={1\over 2}\,p^{-2}_{{}_\Phi}
\left(\Phi+{1\over \Phi}-2\right)
\la{V}
\een
is {\em the
dilatonic potential} \cite{F02}.

\subsection{First Normal Form of the Dynamical Equations}

Introducing a new regularizing variable $\tau$ by $dt=H d\tau$ and
the notations $x_1:=H^2 \geq 0$, $x_2:=\Phi > 0$, $x_3:=\ln a$
(where $H=\dot a/a$ is the Hubble parameter) we can rewrite the
basic system (\ref{DERWU}) in standard normal form: \ben {\frac d
{d\tau}}{\bf x}= {\bf f(x)} \,, \la{Norm} \een where ${\bf x},
{\bf f}\in {\cal R}^{(3)}$ are three-dimensional vector-columns
with components $\{x_1,x_2,x_3\}$ and $\{f_1,f_2,f_3\}$,
respectively, and \ben f_1= 4 x_1
\left(Z_1(x_2)-x_1-ke^{-2x_3}/2\right),
\hskip 1.1truecm \nonumber \\
f_2=x_2\left( Z_2(x_2) -x_1-  k e^{-2x_3}\right),\hskip 1.65truecm   \nonumber \\
f_3=x_1.\hskip 5.35truecm
\la{f}
\een
Here the quantities
\ben
Z_1(\Phi)={\frac 1 6}U{\!,_\Phi}(\Phi)=
Z_2(\Phi)+{\frac 1 4}V_{\!,_\Phi}/\Phi,
\nonumber \\
Z_2(\Phi)={\frac 1 3}U(\Phi)/\Phi= {\frac 1 3}\Phi+{\frac 1
3}W/\Phi. \hskip .4truecm \la{Z1Z2} \een are regarded as functions
of $x_2=\Phi$.

Now one can derive another important relation of a
pseudo-energetic type by using Eq.~(\ref{EqPhi}). The qualitative
dynamics of its solutions is determined by the function \ben
\eta={1\over 2}\dot\Phi^2+V(\Phi). \la{eta} \een This function
plays the role of a (non conserved) energy-like function for the
dilaton field $\Phi$ in the 4D-DG-RW Universe and obeys the
equation \ben {d\over {d\tau}} \eta = - 3 \left({d\over
{d\tau}}\Phi\right)^2 \leq 0. \la{Eq_eta} \een

As seen from this equation, the quantity $\eta$ defines a Lyapunov
function in a generalized sense in our model of Universe, filled
only with vacuum energy. In this case the parameter $\eta(\tau)$
is a monotonically decreasing function of the time $\tau$. One can
show, that the same relations are valid for the case of {\em ultra
relativistic matter}, when the trace of energy-momentum tensor equals
zero and the function $\eta$ is a Lyapunov function for the
corresponding generalization of the normal system of ordinary
differential equations (\ref{Norm}) \cite{F02}.

\begin{figure}[htbp]
\vspace{6.5truecm} \includegraphics{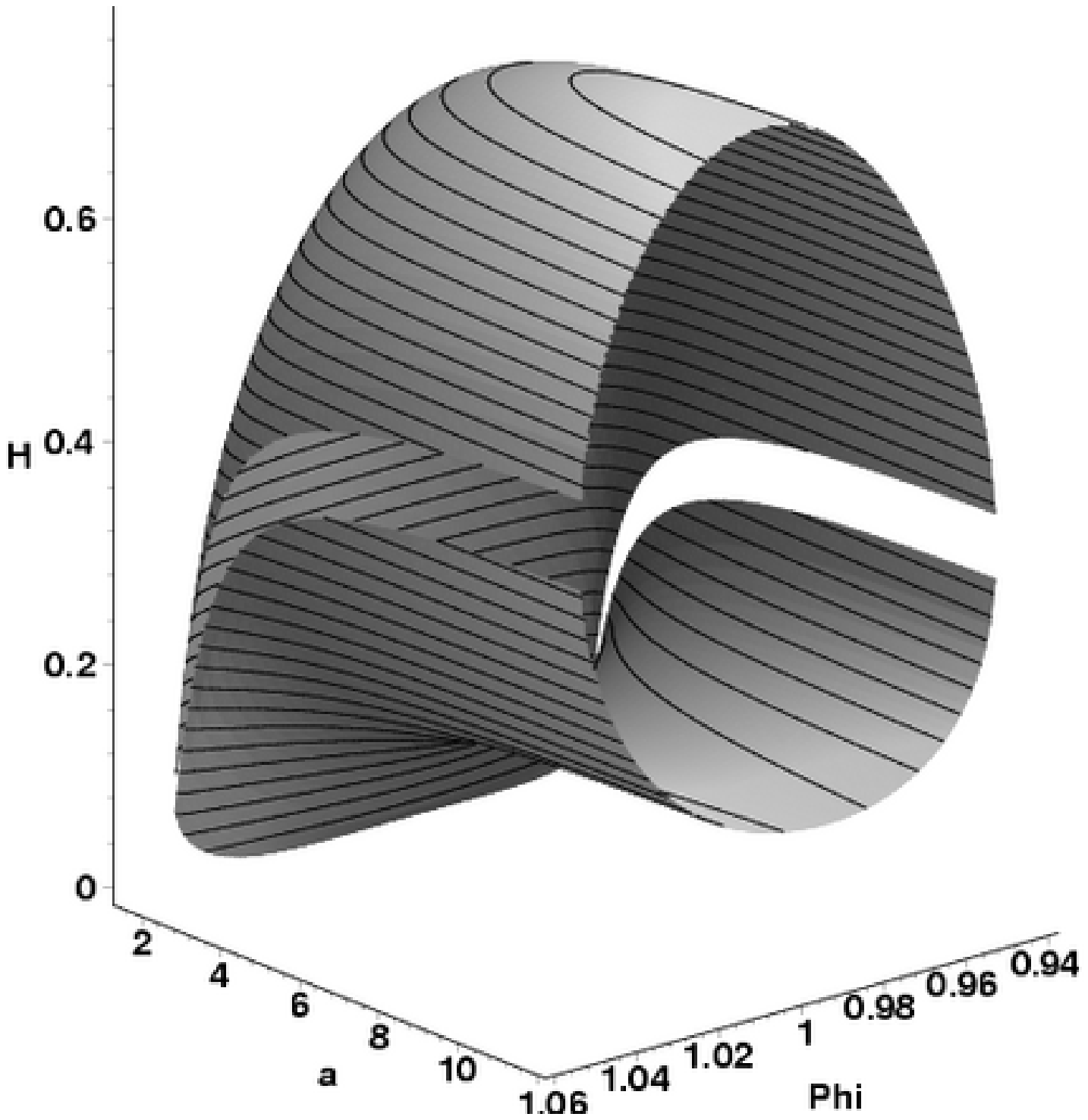}
\caption{\hskip 0.2truecm Level surface
        of the Lyapunov function $\eta$ -- Eq.~(\ref{eta})
        for the potential $V(\Phi)$ -- Eq.~(\ref{Vsimplest}) and $k=+1$.
    \hskip 1truecm}
    \label{FigL1p}
\end{figure}

\begin{figure}[htbp]
\vspace{6.5truecm} \includegraphics{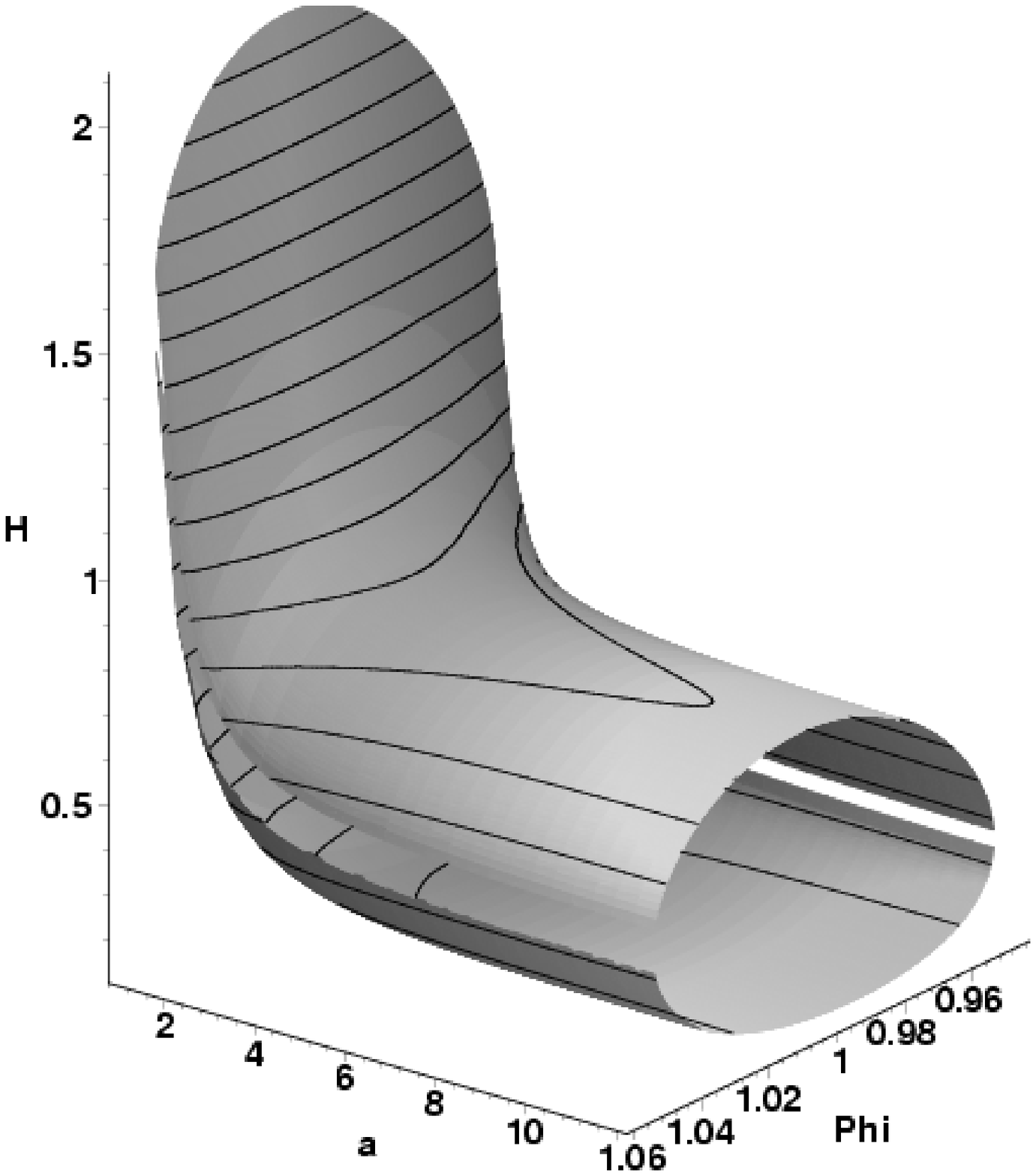}
\caption{\hskip 0.2truecm Level surface
        of the Lyapunov function $\eta$ -- Eq.~(\ref{eta})
        for the potential $V(\Phi)$ -- Eq.~(\ref{Vsimplest}) and $k=-1$.
    \hskip 1truecm}
    \label{FigL1m}
\end{figure}

All solutions to the system (\ref{Norm}) must cross the surfaces
of level $\eta=const$ of this function and go into its interior.
The form of this two-sheeted surfaces is described explicitly by
the equation: \ben H={1\over\Phi}\sqrt{\eta\!-\!V\!+\!\Phi\Delta
\pm
\sqrt{\left(\eta\!-\!V\right)\left(\eta\!-\!V\!+\!2\Phi\Delta\right)}}.
\la{LyaS} \een Here $\Delta= {1\over3}U-k{\Phi\over{a^2}}$.
A typical geometrical form of a level-surface of the
function $\eta$ {\em in absence of matter} for the simple
potential $V=V(\Phi)$ (\ref{V}) is shown in Fig.~\ref{FigL1p} --
for $k=1$, and  in Fig.~\ref{FigL1m} -- for $k=-1$. For $k=0$ the
function (\ref{LyaS}) does not depend on the scale factor $a$ and
the corresponding surface is a simple cylinder.

One can see two main specific features of the behavior of this
surface:

1) At small values of the RW scale factor $a\sim 0$ the form of
the Lyapunov surface depends strongly on the sign of 3-space
scalar curvature: For $k=+1$ we have a semi-infinite curved
cylinder with a closed bottom, which goes down to values $H=0$ in
''vertical'' direction. In contrast, for $k=-1$ we have on both
sides an infinite curved cylinder which goes up to values
$H=\infty$ in ''vertical'' direction. At the end, for $k=0$ we
have a simple straight ''horizontal'' cylinder.

2) For all values $k=0,\pm 1$ and $a\to\infty$ we have infinitely
long ``horizontal'' tube along the $a$-axes. The solutions that
enter the inner part of the $\{a,\Phi,H\}$-space through the
Lyapunov surface, in this area go to de Sitter asymptotic regime
winding around the axis of the tube and approaching it for $t \to
\infty$.

The analytical behavior of the solutions in the completely
different regimes at $a\sim 0$, and at $a \gg 1$, as well as the
transition of given solution to asymptotically de Sitter regime
will be described below in more detail.

\vskip .5truecm

\subsection{Second Normal Form of the Dynamical Equations\\}

We need one more normal form of the dynamical equations to study
the behavior of their solutions for $x_1,x_2\to \infty$. The
standard techniques  for investigation of the solutions in this
limit show that the infinite point is a complex singular point,
and one has to use the so called $\sigma$-process  \cite{Arnold}
to split this singular point into elementary ones. This dictates
the following change of variables: \ben x_1={\frac 3
{16}}p_{{}_\Phi}^{-2}z^{-4}g^{-4},\,\,\, x_2=g^{-1},\,\,\,t={\frac
4 {\sqrt{3}}}\,p_{{}_\Phi}\Theta\,, \la{x1x1zg} \een which
transforms the equations (\ref{Norm}) with right hand sides
(\ref{f}) into a new system: \ben g^\prime=g\,{\cal D}, \,\,\,
z^\prime=z{\cal Z}, \,\,\, \Theta^\prime=z^2 g^2 \la{Norm2} \een
where the prime denotes differentiation with respect to the
variable $x_3$ and \ben {\cal D}(g,z,a;p_{{}_\Phi}^2,k) &=&
1-{\frac 1 3}z^4 g^5 w(g) \nonumber \\
&-& \left(\!{\frac{4p_{{}_\Phi}}3}\!\right)^{\!2}\!\!g^3 z^4\!
\left(\!1\!-\!3g{\frac k {a^2}}\!\right)
\!     \nonumber \\
&=&{\cal D}_0(g,z)\!+\!{\cal O}_2(p_{{}_\Phi}^2)
\nonumber \\
{\cal Z}(g,z,a;p_{{}_\Phi}^2,k) &=&
{\frac 1 4}z^4g^7 v{,_g}(g)   \nonumber \\
&-& \!{\frac 3 2}\left(\!{\frac {4p_{{}_\Phi}} 3}\!\right)^{\!2}\!g^4 z^4\!
\left({\frac k {a^2}}\!\right)
\!  \nonumber \\
&=&\!{\cal Z}_0(g,z)\!+\!{\cal O}_2(p_{{}_\Phi}^2) \,.
\la{D}
\een
Here we have introduced the functions
$v(g)\!=\!{\frac {16} 3}p_{{}_\Phi}^{2}\!V({\frac 1 g})$ and
$w(g)\!=\!{\frac {16} 3}p_{{}_\Phi}^{2}\!W({\frac 1 g})$
{\em which do not depend on the parameter} $p_{{}_\Phi}$ and
are simply related: $w=3/2 g^{-2}\int_1^g g^3 dv$.
The functions ${\cal D}$ and ${\cal Z}$
are defined according to the formulae
\ben
{\cal D}= {\frac {d(\ln g)}{d(\ln a)}}\,,\quad
{\cal Z}= {\frac {d(\ln z)}{d(\ln a)}}\,.
\la{Ddef}
\een

The representation (\ref{D}) shows that one can
develop a simple perturbation theory for the highly
nonlinear system (\ref{Norm2}) using the extremely
small parameter $p_{{}_\Phi}^{2}$ as a
perturbation parameter.

From the dynamical equations (\ref{Norm}), (\ref{f}) and
(\ref{Norm2}), one easily obtains the following contour-integral
representation for the number of e-folds ${\cal N}$ and for the
elapsed time  $\Delta t_{\cal N}= {\frac 4
{\sqrt{3}}}\,p_{{}_\Phi}\Theta_{\cal N}$: \ben {\cal
N}(p_{{}_\Phi}^2,k)\!=\! \int_{\tau_{in}}^{\tau_{fin}} \limits
H^2(\tau) d\tau\!=\!\! \int_{{\cal C}_{in}^{fin}}\limits\! {\frac
{dg/g}{{\cal D}(g,z,a;p_{{}_\Phi}^2,k)}}, \la{N} \een and \ben
\Theta_{\cal_N}(p_{{}_\Phi}^2,k)= \int_{{\cal
C}_{in}^{fin}}\limits\! {\frac {z^2 g\,dg}{ {\cal
D}(g,z,a;p_{{}_\Phi}^2,k)} }. \la{TN} \een Here, a start from some
initial (in) state of the 4D-DG-RW Universe, followed by a motion
along the contour ${\cal C}_{in}^{fin}$ (determined by
corresponding solution of system (\ref{Norm2})), and an end at
some final (fin) state, are assumed.

\section{General Properties of the Solutions
            in the 4D-DG RW Universe}

\subsubsection{Properties of the Solutions in a Vicinity of dSV}

Let us first consider the simplest case when
$k=0$. The system (\ref{Norm})
in this case splits into a single equation for
$x_3$, which is solved by the {\em monotonic} function
$x_3(\tau)=x_3^0+\int_{\tau_0}^\tau x_1(\tau)d\tau$,
and the independent of $x_3$ system
\ben
{\frac d {d\tau}}{ x_1}=4 x_1\left(Z_1(x_2)-x_1\right),
\nonumber \\
{\frac d {d\tau}}{ x_2}=\,\,x_2\left( Z_2(x_2) -x_1\right).
\hskip 0.truecm
\la{kE0}
\een

Now it is clear that the curves $\hat x_1= Z_1(\hat x_2)$ and
$\check x_1= Z_2(\check x_2)$ are the zero-isoclinic lines for the
solutions of (\ref{kE0}). These curves describe the points of
local extrema of the functions $x_1(\tau)$ and $x_2(\tau)$,
respectively, in the domain $x_{1,2}>0$. Because of the condition
$U(\Phi)>0$  and the existence of a unique minimum of the
cosmological potential, these lines have a unique intersection
point -- a de Sitter vacuum (dSV) state \cite{F02} with $\bar
x_1=1/3$, $\bar x_2=1$. This singular point represents the
standard de Sitter solution, which in usual variables reads \ben
\bar H = 1/\sqrt{3},\,\,\,\bar \Phi=1,\,\,\, a(t)=a_0
\exp(t/\sqrt{3}). \la{dSS} \een

\begin{figure}[htbp]
\vspace{6.truecm} \includegraphics{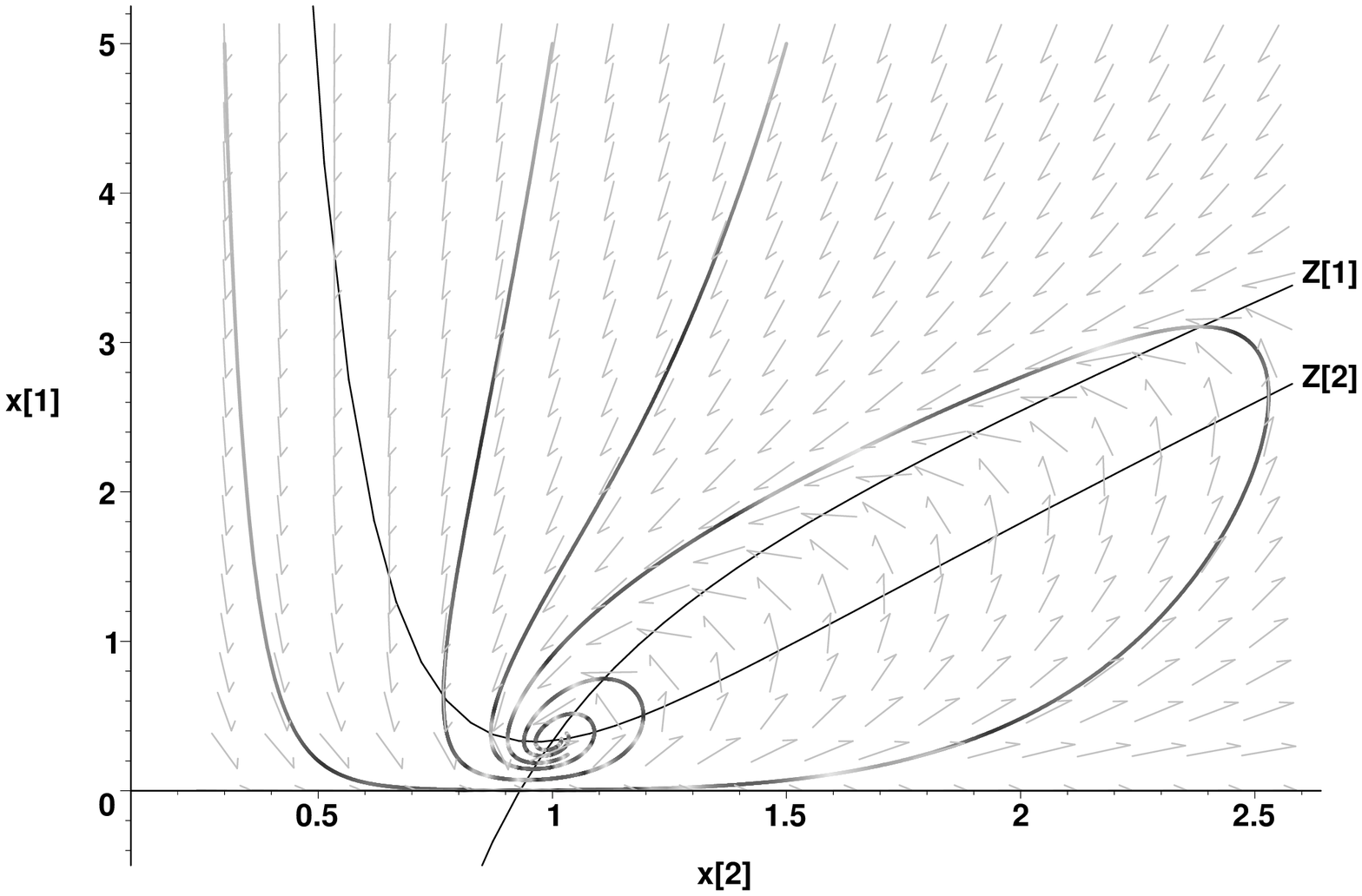}
\caption{\hskip 0.2truecm A typical phase portrait of the system
(\ref{kE0}).
        The parts of solutions with the same color are covered
        by the Universe for equal $\tau$-intervals.
    \hskip 1truecm}
    \label{Fig1}
\end{figure}

One can see the typical behavior of the solutions
of the system (\ref{kE0}) in the domain $x_{1,2}>0$,
together with the curves
$Z[1] := \{\hat x_1= Z_1(\hat x_2)\}$,
and
$Z[2]:=\{\check x_1= Z_2(\check x_2)\}$,
in Fig.\ref{Fig1}, where the corresponding phase portrait
is shown for the case of the potentials (\ref{Vsimplest})
and $p_{{}_\Phi}=1/4$.

Consider the solutions of the system (\ref{kE0}) of the form
$x_{1,2}(\tau)=\bar x_{1,2}+\delta x_{1,2}(\tau)$ that are close
to dSV, i.e., with $|\delta x_{1,2}(\tau)|\ll 1$. Using the
relations
\ben V(1)=0,\,\,\,V{\!,_\Phi}(1)=0,\,\,\,
V{\!,_{\Phi\Phi}}(1)=p^{-2}_{{}_\Phi};
\hskip 1.truecm\nonumber\\
U(1)=1,\,\,\,U{\!,_\Phi}(1)=2,\,\,\, U{\!,_{\Phi\Phi}}(1)=
2\left(1+{3\over4}p^{-2}_{{}_\Phi}\right) \,, \la{dSVcommon} \een
which describe the general properties of de Sitter vacuum state
(with $\bar \Phi =1$) in 4D-DG \cite{F02}, one obtains in linear
approximation \ben \delta x_1(t)= \delta
x_1^0\,\exp\!\left({-{\frac {\sqrt{3}} 2}t}\right)
\cos\left(\omega_{{}_\Phi} t\right),
\hskip 1.4truecm \nonumber \\
\delta x_2(t)\!=\!{\frac 3 2}
{\frac { p_{{}_\Phi}\delta x_1^0}
{\sqrt{ p_{{}_\Phi}^2\!+\!{\frac 3 4}}}}
\exp\!\left(\!{-{\frac {\sqrt{3}} 2}t}\!\right)\!
\cos\left(\omega_{{}_\Phi}t\!-\!\psi_{{}_\Phi}\right)
\nonumber \\
\delta x_3(t)\!=\!\delta x_3^0\!+
\!\sqrt{3} p_{{}_\Phi}\delta x_1^0
\exp\!\left(\!{-{\frac {\sqrt{3}} 2}t}\!\right)\!\times
\hskip 1.2truecm \nonumber \\
\bigl(\cos\left(\omega_{{}_\Phi}t\!-\!\psi_a\right)-
\cos\left(\psi_a\right)
\bigr)\,,
\la{dxt}
\een
where
\ben
\omega_{{}_\Phi}=\sqrt{p_{{}_\Phi}^{-2}-3/4}\,,
\hskip 1.5truecm \nonumber \\
\tan\psi_{{}_\Phi}={\frac {2\sqrt{3}} 5}\,\omega_{{}_\Phi},
\,\,\,
\tan\psi_a=-{\frac {2\sqrt{3}} 3}\,\omega_{{}_\Phi},
\la{omega}
\een
$\delta x_1^0=\delta x_1(0)$ is the small
initial amplitude of $\delta x_1(t)$,
and the solution for the deviation $\delta x_3(t)$
has been added
for completeness and for later use.

The frequency $\omega_{{}_\Phi}$ is a real and
positive number if $p_{{}_\Phi} <2/\sqrt{3}$.
According to the estimate $p_{{}_\Phi}< 10^{-30}$, in 4D-DG we have
$\omega_{{}_\Phi} \geq 10^{30}$ in cosmological units,
or $\omega_{{}_\Phi} \geq  10^3\, {\rm GHz}$ in usual units,
and $\psi_{{}_\Phi}\approx \pi/2\approx -\psi_a$.

Thus, we see that the de Sitter vacuum is a stable focus in
the phase portrait of the system (\ref{kE0}).
For $k=0$  all solutions of
this system that lie in a small enough vicinity
of dSV oscillate with an ultra-high frequency
$\omega_{{}_\Phi}$ (\ref{omega}),
and approach dSV in the limit $t \to \infty$.

One can  generalize this statement for the
case of arbitrary $k=0,\pm1$ \cite{F02}.
This turns out to be true even in
the presence of nonzero matter energy-density terms.
Then we see that one important general prediction
for the 4D-DG-RW Universe is the existence of ultra-high
dilatonic oscillations with frequency $\omega_\Phi$ (\ref{omega})
in 3-spaces with any curvature and in the presence
of any kind of normal matter \cite{F02}.

If $p_{{}_\Phi} \geq 2/\sqrt{3}$, i.e., if $m_{{}_\Phi}\sim 10
^{-33}\, {\rm eV}$ (as in inflation models with a slow-rolling
scalar field and in quintessence models \cite{C, inflation, Q}),
the above ultra-high dilatonic oscillations do not exist in the
4D-DG-RW Universe. In this case, the frequency $\omega_{{}_\Phi}$
becomes imaginary and, instead of a stable focus, we have an
unstable saddle point in the phase portrait of the system
(\ref{kE0}). Such a situation was considered first in
\cite{Starobinsky80} in a different model of nonlinear gravity
based on a quadratic with respect to scalar curvature $R$
gravitational Lagrangian, but with some additional terms that
originate from quantum fluctuations in curved space-time
\cite{QNLG}. These additional terms are related to the Weyl
conformal curvature and therefore vanish in the case of RW metric,
but yield an essentially different theory in other cases.
Therefore, for RW Universe the model described in
\cite{Starobinsky80} is equivalent to 4D-DG with a quadratic in
the field $\Phi$ cosmological potential. Such a potential is
non-physical, because it admits negative values of the dilaton
field $\Phi$, i.e., of the gravitational constant.

An immediate consequence of the above consideration is the
existence of ultra-high frequency oscillations of the effective
gravitational factor $G_{eff}= G_{{}_N}/\Phi$, accompanied with an
extremely slow exponential decrease of its amplitude $\sim
\exp\left(-{\frac 1 2}\sqrt{3\Lambda^{obs}}\,ct\right)$ (in usual
units). One obtains $\bar H^2/ H^2_0=\Omega_\Lambda$ and $\delta
H^2_0 = {\frac 1 3}{\frac {1- \Omega_\Lambda}{\Omega_\Lambda}}$.
Hence, at the present epoch with $\Omega_\Lambda \approx {\frac 2
3}$, we have $\delta x_1^0\approx {\frac 1 6}$. Then the second
equation of the system (\ref{dxt}) gives \ben g(t) \approx 1 -
p_{{}_\Phi} {\frac {\sqrt{3}} 2} \exp\!\left({-{\frac {\sqrt{3}}
2}t}\right) \cos\left(\omega_{{}_\Phi}t-\psi_{{}_\Phi}\right)\,,
\la{Gvar} \een where we have introduced a dimensionless
gravitational factor, $g(t)={ {G_{eff}(t)}/{G_{{}_N}}}=1/\Phi(t)$.

Because of the extremely small amplitude
$\!p_{{}_\Phi}\!\leq\!10^{-30}$,
these variations are beyond the possibilities
of present-day experimental techniques.

In contrast, the oscillations of the Hubble parameter
$H$ have a relatively big amplitude
$\delta H_0=\sqrt{\delta x_0}\approx 0.4$, and the
same huge frequency $\omega_{{}_\Phi}$,
as the oscillations of gravitational factor.
It is very interesting to find possible
observational consequences of such a phenomenon.

High frequency oscillations of the effective gravitational
factor were considered first in the context
of Brans-Dicke field with a BD parameter $\omega >1$
in \cite{Steinhardt}. These oscillations were induced
by an independent inflation field, but the analysis of the existing
astrophysical and cosmological limits on the oscillations of
$G_{eff}(t)$ is  applicable to our 4D-DG model as well.
The conclusion in  \cite{Steinhardt} is that
the oscillations in the considered frequency-amplitude range,
being proportional to $\dot g/(gH)$,
do not affect the Earth-surface laboratory measurements,
Solar System gravitational experiments,
stellar evolution, and nucleosynthesis, but can produce
significant cosmological effects because the frequency
is very large and the Hubble parameter is small
(in usual units). It can be seen explicitly from Eq.~(\ref{dxt})
that this is precisely what happens in 4D-DG,
although in it the oscillations are self-induced.

As stressed in \cite{Steinhardt}, despite the fact
that the variations of the type (\ref{Gvar}) have
extremely small amplitudes, they can produce significant
cosmological effects because of the nonlinear
character of gravity. In the case of free of matter Universe the
4D-DG version of the formula, analogous to the one
in the first of references \cite{Steinhardt}, is
\ben
H={\frac 1 2}{\frac{\dot g} g }\pm
\sqrt{{\frac 1 4} \left({\frac {\dot g} g}\right)^2
+{\frac 1 3}gU -{\frac k{a^2}}}\,.
\la{H_nlin}
\een
Being a direct consequence of Eq.~(\ref{DERWU}),
this formula shows that, after averaging of
the oscillations, the term ${{\dot g}/g}$
has a non-vanishing contribution
because it enters the Hubble parameter
(\ref{H_nlin}) in a nonlinear manner.
A more detailed mathematical treatment of this new phenomenon
in 4D-DG is needed to derive reliable conclusions.
The standard averaging techniques for differential equations
with fast oscillating solutions and slowly developing modes
seem to be the most natural mathematical method for this purpose,
but the applications of these techniques
to 4D-DG lie beyond the scope of the present article.

\section{Inflation in 4D-DG-RW Universe}

Having in mind that: 1) the essence of the inflation is a fast and
huge re-scaling of the Universe \cite{C, inflation}, and 2) the
dilaton is the scalar field determining the scales in
Universe, it seems natural to relate these two fundamental
physical notions instead of inventing some specific ``inflation
field''. In this section we show that our 4D-DG model indeed
offers such a possibility.

\subsection{ The Phase-Space Domain of Inflation\\}

As seen from the phase portraits in Fig.~\ref{Fig1}--\ref{Fig2},
for values $H\geq H_{crit}$ and $\Phi\geq \Phi_{crit}$,
%i.e. above some critical values $\{H_{crit},\Phi_{crit}\}$,
the ultra-high-frequency-oscillations do not exist. The evolution
of the Universe in this domain of the phase space of the system
(\ref{DERWU}) reduces to some kind of monotonic expansion,
according to the equation ${\frac d {d\tau}} x_3=H^2>0$. We call
this expansion an {\em inflation}. As we shall see, it indeed has
all needed properties to be considered as an inflation phenomenon.

The transition from inflation to high-frequency
oscillations is a nonlinear phenomenon, and
we will describe it in the present article very
briefly. Here our goal is to have some
approximate criteria for determining the end of
the inflation. It is needed for evaluation of the basic
quantities that describe the inflation.

As seen from Eq.~(\ref{dxt}), the amplitude, $\delta \Phi_0$,
of the oscillations of $\Phi$ is extremely small
compared with the amplitude $\delta H^2$
of the oscillations of $H^2$:\,\,
$\delta\Phi_0 \alt {\frac 3 2}p_{{}_\Phi}/
\sqrt{p_{{}_\Phi}^2+{\frac 3 4}}\,\delta H_0^2$.
An obvious estimate for the amplitude
$\delta H_0^2$ is $\delta H_0^2\leq 1/3\,(=\bar H^2)$.
Then, for $p_{{}_\Phi}\leq 2/\sqrt{3}$ (which is the
condition for existence of oscillations), we obtain
$\delta \Phi_0 \alt\delta H_0^2\alt 1/3$.
The last estimate is indeed very crude for
the physical model at hand, in
which $p_{{}_\Phi}\leq 10^{-30}$.
This consideration produces the constraint
$H_{crit}\alt\sqrt{2/3}$ and $\Phi_{crit}\alt 4/3$,
but, taking into account the extremely small value
of $p_{{}_\Phi}$, we will use for simplicity the
very crude estimate $H_{crit}, \Phi_{crit}\sim 1$.

Now it becomes clear that the study of the inflation
requires considering big values of the variables $x_{1,2}$,
i.e., using the second normal form (\ref{Norm2})
of the dynamical equations.

\subsection{ The Behavior of the Solutions Near the Beginning\\}

\subsubsection{The Case  $k=0$}

Let us consider first the case $k=0$. From Eq.~(\ref{Norm2})  one
obtains the simple first order equation \ben
{\frac{dz}{dg}}={\frac {z^5 g^6}{4}} {\frac{v_{,_g}(g)}{{\cal
D}(g,z;p_{{}_\Phi}^2)}} \la{zg} \een with \ben {\cal
D}(g,z;p_{{}_\Phi}^2)=1-{\frac 1 3}z^4g^5 u(g)=
\hskip 3.truecm\nonumber \\
1\!-\!{\frac 1 3}z^4 g^5 w(g)\!-\!
\left(\!{\frac{4p_{{}_\Phi}}3}\!\right)^{\!2}\!\!g^3 z^4\!=\!
{\cal D}_0(g,z)\!+\!{\cal O}_2(p_{{}_\Phi}^2),\,\,\,\,
\la{D00}
\een
where $u(g)={\frac{16}{3}}p_{{}_\Phi}^2 U(1/g)$.

One can easily prove that the solutions of Eq.~(\ref{kE0}) have
important general properties for small values of $g$. For the
potentials (\ref{V}) we obtain
\ben {\cal Z}_0(g,z)=
-{\frac{2z^4}{3}} g^5\left(1 - g^{2}\right) \la{Znu} \een
and
\ben
{\cal D}_0(g,z)\!=\! 1\!-\!{\frac {z^4}{3}}g^3
\left(1-g^2\right)^2. \la{Dnu} \een

Hence, $\lim_{g\to 0}{\cal D}_0(g,z)=1$.
Then, taking into account the leading terms, we obtain from
Eq.~(\ref{Norm2}), (\ref{Ddef}), (\ref{TN}), (\ref{zg}), and
(\ref{D00}) the following results in the limit $g\to 0$:
\ben
a(g)\sim g,\hskip 5.2truecm
\nonumber\\
z(g)\!=\!z_0\!- {\frac{2}{15}}\,z_0^5 g^{5}
+{\cal O}(g^6),\hskip 2.3truecm\nonumber\\
t(g)\!=\!{\frac{2p_{{}_\Phi}}{\sqrt{3}}}z_0^2 g^2 \!+\!{\cal
O}(g^3).\hskip 3.05truecm
\la{azt_g}
\een

When solved with respect to the time $t$, these formulae give
\ben
a(t)\sim\left({\frac t { p_{{}_\Phi}}}\right)^{1/2}\!\!
,\,\,\,g(t)={\frac 1 {z_0}} \left({ \frac {\sqrt{3}} {2
p_{{}_\Phi}} } t\right)^{1/2}\!\!
+{\cal O}(t),\nonumber\\
z(t)=z_0-{\frac {2} {15}}\left(\!{\frac{\sqrt{3}} {2
p_{{}_\Phi}}}t\!\right)^{5/2}+ {\cal O}(t^{3}).\,\,\, \la{agz_t}
\een

\subsubsection{The Case  $k=\pm 1$}

Now, the equations (\ref{Norm2}) do not split into a two-dimensional
subsystem and an independent differential equation and must be
considered as a true three-dimensional system. The expansion of
its solutions in a Taylor series with respect to the variable $a$ is
\ben
g(a)=g_1\, a+{\cal O}(a^2),\hskip 2.1truecm \nonumber \\
z(a)=z_0-{\frac 4 3 }\,k\, p_{{}_\Phi}^2 \,g_1^4 \,a^2
+ {\cal O}(a^3),\,\,\,\nonumber \\
t(a)={\frac 2 {\sqrt{3}} }\, p_{{}_\Phi}\, z_0^2\, g_1^2\, a^2+
{\cal O}(a^3),\hskip .6truecm
\la{gzt_a}
\een
with an arbitrary
constants $g_1$ and $z_0$. When solved with respect to the time $t$,
these formulae give
\ben
a(t)=z_0^{-1}g_1^{-1}\left({\frac{\sqrt{3}}{2p_{{}_\Phi}}}t
\right)^{1/2} +{\cal O}(t)\hskip .5truecm \nonumber \\
g(t)=z_0^{-1}\left({\frac{\sqrt{3}}{2p_{{}_\Phi}}}t \right)^{1/2}
+{\cal O}(t)\hskip 1truecm \nonumber \\
z(t)=z_0 -k \left({\frac {g_1}{z_0}}\right)^2{\frac
{2p_{{}_\Phi}}{\sqrt{3}}}t+{\cal O}(t^{3/2}).
 \la{agz_t_pm}\een
We see, that the value of the sign $k=0,\pm 1$ of the 3-space
curvature influences only the form of the functions $z(t)$ and
$z(a)$ for small values of the time $t$, and of the scale
parameter $a$, respectively. These differences in the behavior of
the functions $z(t)$ and $z(a)$ for different values of the
parameter $k$ are consistent with the pictures, shown in
Fig.~\ref{FigL1p} and Fig.~\ref{FigL1m}.

\subsubsection{Some General Conclusions}

The above formulae  (\ref{azt_g}), (\ref{agz_t}), (\ref{gzt_a}),
and (\ref{agz_t_pm}) show for any value of parameter $k=0,\pm 1$:

1) the existence of Beginning, i.e., the existence of a time
instant $t=0$ at which $a(0)=0$;

2) the asymptotic freedom of gravity, i.e., the zero value of
gravity at the Beginning: $g(0)=0$;

3) the finiteness of the time interval needed for reaching nonzero
values of $a$ and $g$, starting from the Beginning;

4) the high precision constancy of $z(t)$ at the Beginning for
$k=0$ and  the dependence of linear term in the Taylor series
expansion of the function $z(t)$ for small values of cosmic time
$t$ on the sign $k=\pm 1$ of the 3-space curvature.

5) the behavior of the RW scale factor $a(t)$ for small $t$,
similar to its behavior in GR in the presence of radiation. Hence,
one can conclude that at the Beginning the dilaton plays a role,
similar to the role of a radiation.

The asymptotic freedom of gravity at the Beginning in 4D-DG-RW
Universe is consistent with the qualitative results in
\cite{Reuter}. It leads to some sort of an initial power-law
expansion. The number of e-folds ${\cal N}(t) \to -\infty$ as $
t\to +0$ like $\ln t$, since ${\cal N}(t)=\ln a(t)$.

The general conclusion is that the potentials (\ref{Vsimplest}),
(\ref{V}) do not help to overcome the initial singularity problem
in the {\em free of matter} 4D-DG-RW Universe with $k=0,\pm1$.
However, we would like to emphasize that:

1) 4D-DG with the simplest potentials (\ref{Vsimplest}), (\ref{V})
may not be applicable for times smaller then the Planck time
$t_{Pl}\sim 10^{-44}\,{\rm sec}$, because it may be a low-energy
theory, which ignores some essential quantum corrections in
gravi-dilaton sector. (See \cite{F02} for derivation of 4D-DG as a
low-energy limit of the string model.) But one may expect 4D-DG to
be valid after some initial time instant, $t_{in}\sim t_{{}_\Phi}=
\hbar/(m_{{}_\Phi}c^2)$. If $m_{{}_\Phi}\ll M_{Pl}$, we will have
$t_{in}\gg t_{Pl}$ and our results for the case under
consideration may have a physical meaning, leaving open the
initial singularity  problem.

2) Another possibility to overcome the initial singularity problem
is to introduce some kind of matter in the Universe. The
asymptotic freedom of gravity at the Beginning shows that the
gravitational attraction goes to zero for small values of the
scale parameter $a$. Then, the existence of some amount of usual
matter will lead to a positive pressure in the Universe and to
corresponding repealing forces, which will not be compensated by
gravity at small values of the scale factor $a$. The preliminary
numerical investigation showed that for a 4D-DG-RW Universe with
normal matter (dust, perfect fluid, and/or radiation) one indeed has a bouncing solutions $a(t)\neq 0$ without initial
singularities.

\subsection{The Inflation and the Number of E-Folds\\}

Our general analysis of the solutions for 3D-DG RW model of
Universe {\em without matter} shows, both analytically and
numerically, that its qualitative behavior does not depend
essentially on the sign of the 3-space curvature. Therefore, for
simplicity, we consider in what follows only the case $k=0$.

According to  Eq.~(\ref{N}) and Eq.~(\ref{D}) one can represent
the RW scale factor $a(t)$ in the form:
\ben
a(t)=g(t)\exp(\Delta {\cal N}(t)),
\la{a_dN}
\een
where the re-normalized number of e-folds:
\ben
\Delta {\cal N}=
\int_{{\cal C}_{in}^{fin}}\limits\!
{\frac {dg} g}\, {\frac { 1-{\cal D}(g,z,a;p_{{}_\Phi}^2,k) }
{{\cal D}(g,z,a;p_{{}_\Phi}^2,k)}}
\la{dN}
\een
is a finite quantity  in the entire time
interval $t\in [0,\infty)$. Obviously,
$\Delta {\cal N}(t^{(i)})\equiv  {\cal N}(t^{(i)})$
at the special time instants $t^{(i)}$, $i=0,1,...$,
when $g(t)$ reaches its dSV value, i.e.,
when $g(t^{(i)})=1$, for example, in the limit
$t\to\infty$.

\begin{figure}[htbp]
\vspace{6.truecm} \includegraphics{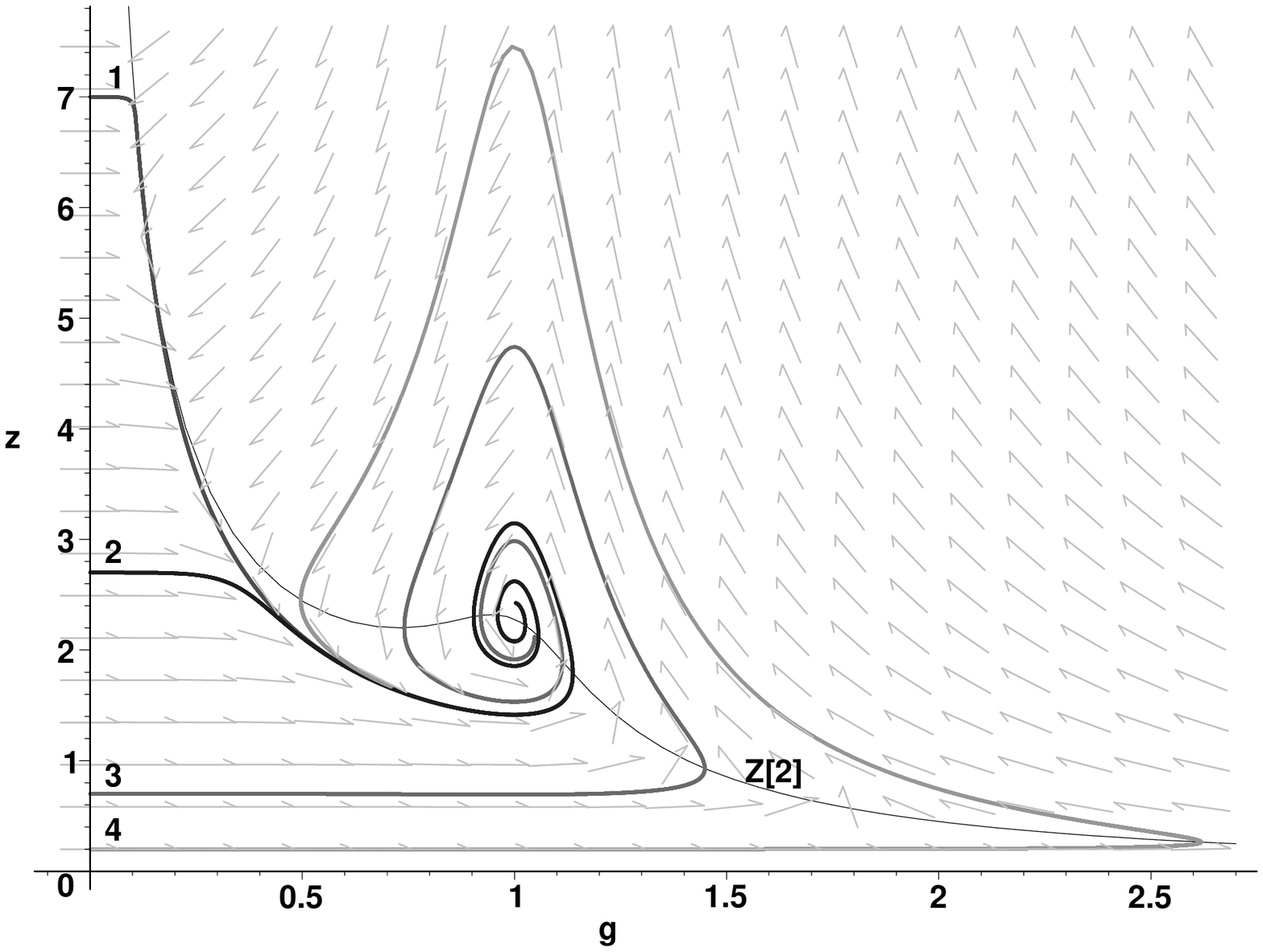}
\caption{\hskip 0.2truecm The phase portrait of Eq.~(\ref{zg}).
The thin line shows the zero line Z[2] of the denominator~${\cal
D}$.
    \hskip 1truecm}
    \label{Fig2}
\end{figure}

\begin{figure}[htbp]
\vspace{5.5truecm} \includegraphics{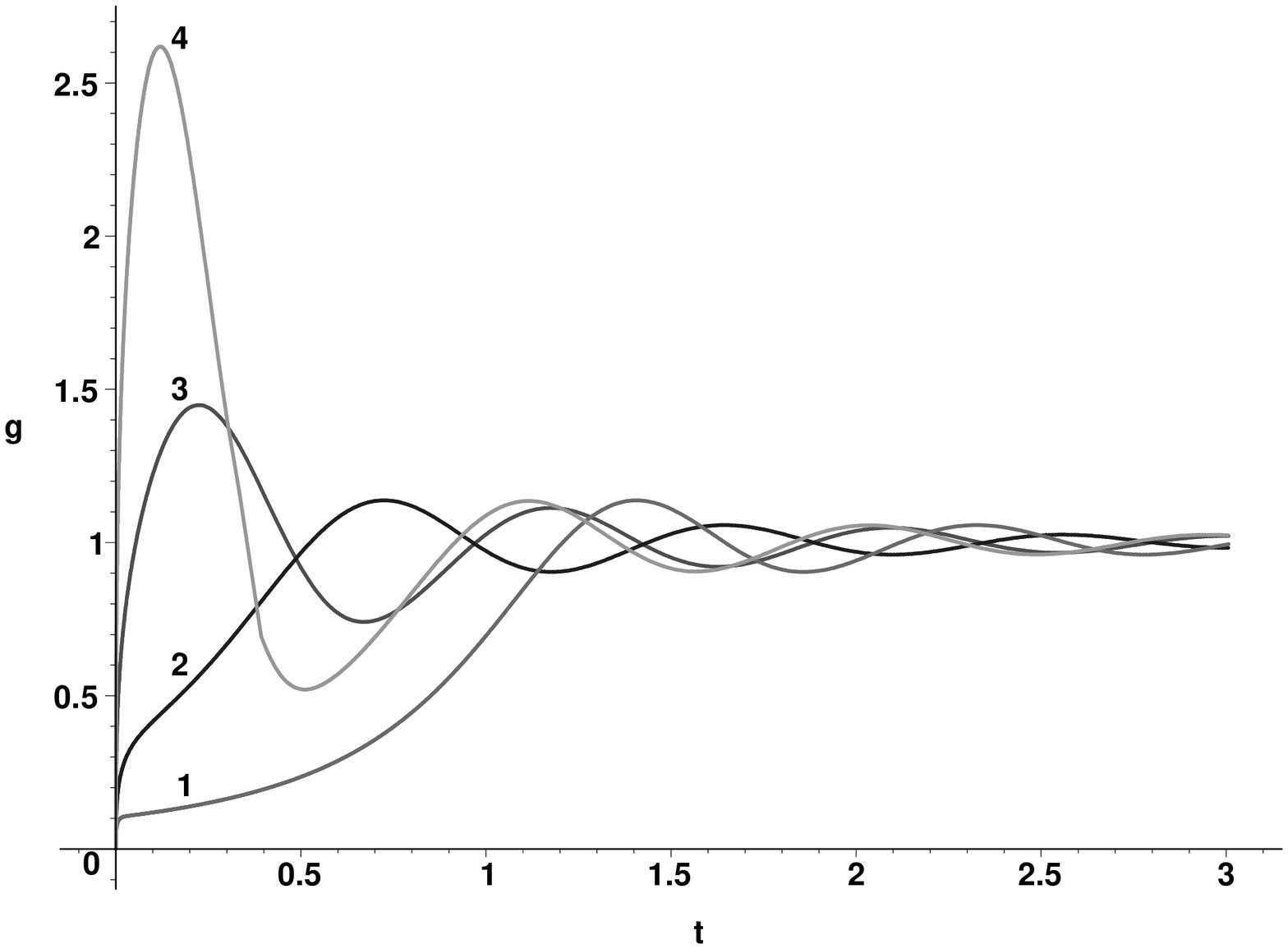}
\caption{\hskip 0.2truecm The dependence of dimensionless gravitational
            factor g on cosmic time t for solutions of Eq.~(\ref{zg}).
    \hskip 1truecm}
    \label{Fig3}
\end{figure}

The phase portrait of Eq.~(\ref{zg}) and the time-dependence of
the dimensionless gravitational factor $g(t)$ for the potentials
(\ref{Vsimplest}), (\ref{V}) are shown in Fig.~\ref{Fig2} and
Fig.~\ref{Fig3}, respectively. From Fig.~\ref{Fig3} we see that
one can define analytically the time of duration of the {\em
initial inflation} $\Delta t_{infl}^{(0)}$ as the time spent by
Universe from the Beginning to the first time instant $t^{(0)}$,
when the gravitational factor $g(t^{(0)})=1$. In addition, we see
in Fig.~\ref{Fig3} that this time interval is finite and has
different values for different solutions of Eq.~(\ref{zg}).

In Fig.~\ref{Fig1} and Fig.~\ref{Fig2} one can see a new specific
feature of 4D-DG: the solutions may enter {\em many times} the
phase-space domain of inflation and the function $g(t)$ oscillates
around its dSV value $\bar g=1$ with a variable period. Between
two successive maxima of $g(t)$, the squared logarithmic
derivative $H^2$ of $a(t)$ has its own maxima, just at the already
defined time instants $t^{(i)}$. In the vicinity of each maximum
of  $H^2$, the function a(t) increases very fast -- like
$\exp({\rm const} \times t^\aleph)$, with ${\rm const}>0$, and
parameter $\aleph \geq 2$. Therefore, we call such inflation,
which is much faster than the usual exponential de Sitter
inflation, a {\em hyper-inflation}. Hence, in 4D-DG, we have some
sort of successive hyper-inflations in the  Universe. For
simplicity, we define the time-duration $\Delta
t_{infl}^{(i)}=t^{(i)}-t^{(i-1)}$, $i=0,1,...$ (by definition
$\Delta t_{infl}^{(0)}=t^{(0)}$) of each of these periods of
hyper-inflation as the time period between two successive dSV
values of the function $g(t)$, although the hyper-inflation itself
takes place only around the maxima of the function $H^2(t)$. The
corresponding number of e-folds is ${\cal N}_{infl}^{(i)}$. It is
clear that ${\cal N}_{infl}^{(i)}$ is a decreasing function of the
number~$i$. The inflation can be considered as cosmologically
significant only if for some short total time period $\Delta
t_{infl}^{total}= \sum_{i=0}^{i_{max}}\Delta t_{infl}^{(i)}$, the
total number of e-folds ${\cal N}_{infl}^{total}=
\sum_{i=0}^{i_{max}}{\cal N}_{infl}^{(i)}$ exceeds some large
enough number ${\cal N}$. It is known that one needs to have
${\cal N} \agt 60$ to be able to explain the special-flatness
problem, the horizon problem, and the large-scales-smoothness
problem in cosmology~\cite{C}.

\begin{figure}[htbp]
\vspace{5.5truecm} \includegraphics{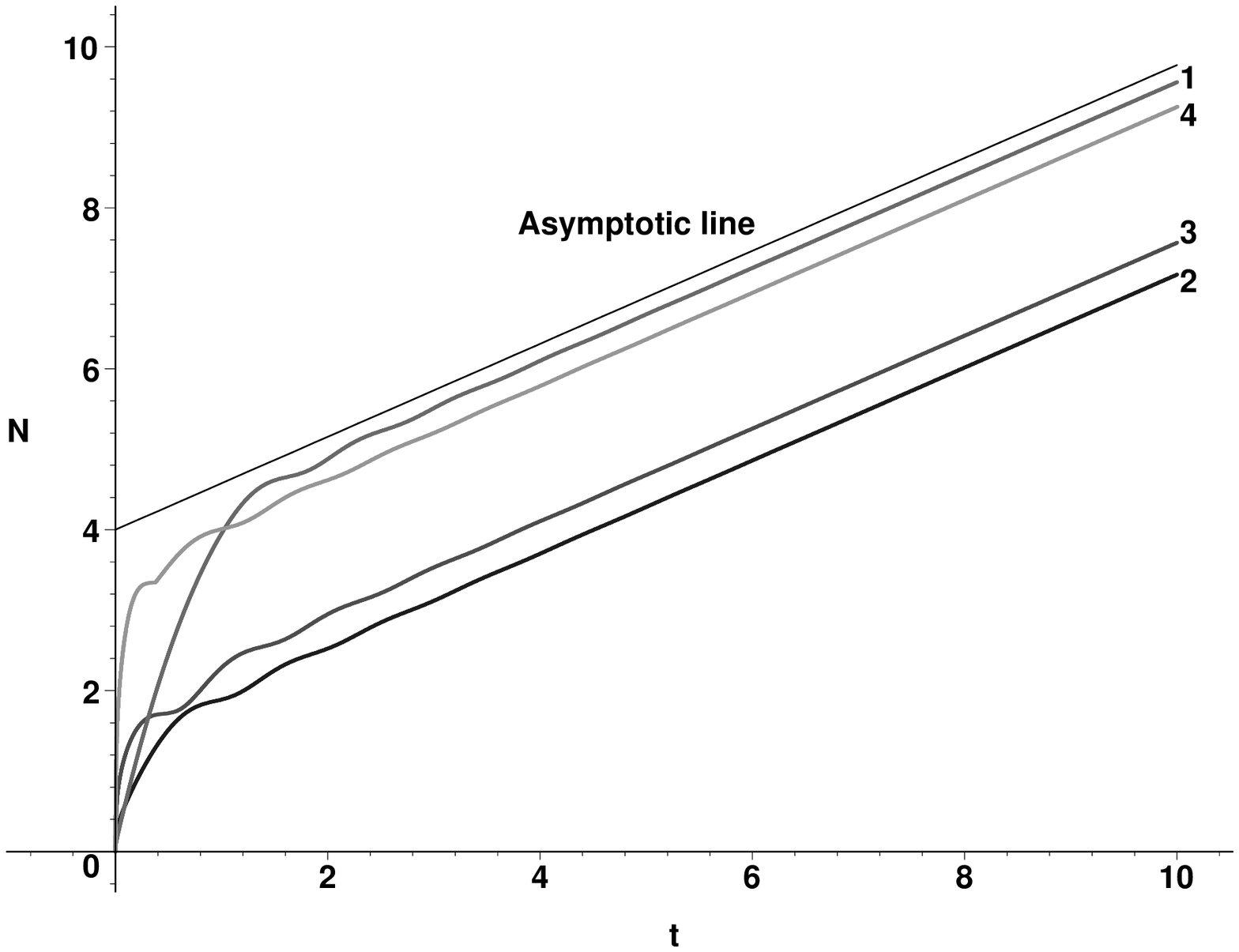}
\caption{\hskip 0.2truecm The dependence of the number of e-folds ${\cal N}$
                      on the cosmic time $t$ for solutions of Eq.~(\ref{zg}).
                      The straight line gives an example of an
                      asymptotic line for some solution.
    \hskip 1truecm}
    \label{Fig4}
\end{figure}

Fig.~\ref{Fig4} illustrates both the inflation and the asymptotic
behavior of the function ${\cal N}(t)$ for $t\to \infty$. The
small oscillations of this function are ``averaged'' by the crude
graphical abilities of the drawing device, and for large values of
time $t$ we actually can see in Fig.~\ref{Fig4} only the limiting
de Sitter regime~(\ref{dSS}), when the averaged function
$\langle{\cal N}(t)\rangle \equiv \langle \Delta{\cal
N}(t)\rangle$. In this regime, for $t \to \infty$, we have an
obvious asymptote of the form \,$\langle{\cal
N}(t)\rangle\sim\bar{\cal N}+\bar H t$ for the averaged with
respect to the dilatonic oscillation (\ref{dxt}) function
$\langle{\cal N}(t)\rangle$. We accept the constant $\bar{\cal N}$
as our {\em final definition of total number of e-folds during
inflation}: \ben \bar{\cal N}=\lim_{t\to\infty}\left(\langle{\cal
N}(t)\rangle
           -\bar H t\right)
            =\lim_{t\to\infty}\left(\langle \Delta{\cal N}(t)\rangle
           -\bar H t\right).\,\,
\la{barN} \een It is clear that this integral characteristic
describes  precisely the number of e-folds {\em due to the true
inflation} in the 4D-DG-RW model, i.e., during the fast initial
expansion of the Universe, as a new physical phenomenon. To obtain
this quantity, one obviously must subtract the asymptotic de
Sitter expansion from the total function $\langle{\cal
N}(t)\rangle$. It can be shown that one has to apply the same
definition to the solutions in the general case of arbitrary values
of the space-curvature parameter $k$ and physically admissible
matter energy densities $\epsilon(a)$, since they have the same
asymptotic behavior. Hence, in all cases the solutions with a
cosmologically significant inflation must have values $\bar{\cal
N} \agt 60$.

As we see in Fig.~\ref{Fig4}, the total number of e-folds
$\bar{\cal N}$ decreases, starting from the solution ``4" (with
$z_0= 0.2$) to the solution ``3" (with $z_0= 0.7$) and to the
solution ``2'' (with $z_0= 2.7$), reaches a minimum ($\approx 1$)
for some initial value $z_0^*$, and then increases for the
solution ``1" (with $z_0=7$), and for solutions with larger values
of $z_0$. Thus, we see that inflation is a typical behavior for
all solutions of Eq.~(\ref{zg}), and that most of them have large
values of $\bar{\cal N}$. (In Fig.~\ref{Fig4}, we show only
solutions with small values of $\bar{\cal N}$ that are close to
its minima.)

The value of the constant $\bar{\cal N}$  depends on the closeness
of the given solution to the zero curve $Z[2]$ of the denominator
of the integrand in the right-hand side of~(\ref{N}). The equation
${\cal D}=0$ can be explicitly solved. Its solution reads
$z=\left({\frac 1 3}g^5u(g)\right)^{-1/4}$. The corresponding
curve is shown both in Fig.~\ref{Fig1} and Fig.~\ref{Fig2}.

The value of the quantity ${\cal N}(t)$ increases essentially each
time when the solution approaches this curve. The corresponding
increment of ${\cal N}(t)$ remains finite, even if the solution
crosses this curve, although in this case the denominator in the
integrals (\ref{N}), or (\ref{dN}), reaches a zero value.

Indeed, taking into account that the points where the solution
crosses the curve $Z[2]$ are extreme points of the function $g(z)$,
and using the expansion $\Delta z= \left(-{\frac 3 8}z{\frac
{gv{,_g}}u}\Delta g\right)^{1/2} +{\cal O}_1(\Delta g)$ in a
vicinity of such a point, we easily obtain
\ben \Delta {\cal N}=
z^{-4}g^{-6}u^{-1} \sqrt{-{\frac{6u}{gv{,_g}}} \Delta g}+{\cal
O}_1(\Delta g),
\nonumber \\
\Delta t={\frac{4p_{{}_\Phi}}{\sqrt{3}}}z^{-2}g^{-4}u^{-1}
\sqrt{-{\frac{6u}{gv{,_g}}} \Delta g}+{\cal O}_1(\Delta g)
\la{DeltaNt}
\een
for potentials $u(g)$ and $v(g)$ of the most
general type. Here the values of the coefficients are taken on the
curve ${\cal D}=0$.

Combined with the previous results, this proves that the time
intervals of inflation $t_{infl}^{(i)}$ are finite for all values
of $i=0,1,\ldots$.

Thus we see that an essential increase in ${\cal N}(t)$ is
accumulated when the solution becomes close and parallel to the
curve $Z[2]$. This is possible both for big values of $g$ and
small values of $z$ (the solution ``4"), or for small values of
$g$ and big values of $z$ (the solution ``1"). All this results in
a big value of $\bar{\cal N}$. Only a small fraction of the
solutions (like the solutions ``2'' and ``3'') stay all the time
relatively far from the line ${\cal D}=0$, and therefore acquire
relatively small number of e-folds, $1\lesssim\bar{\cal N}\lesssim
2-3$.
\begin{figure}[htbp]
\vspace{5.5truecm} \includegraphics{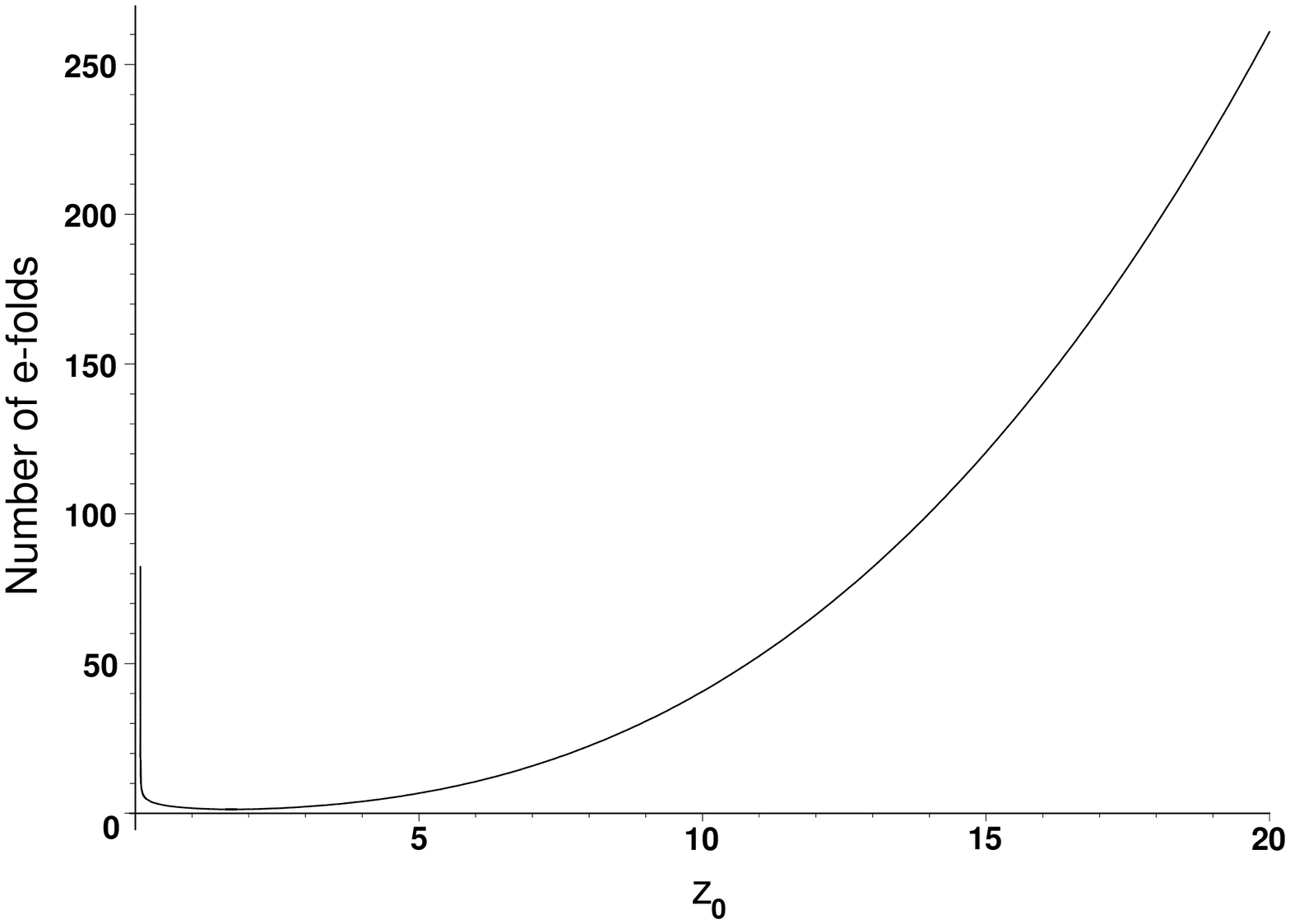} \caption{\hskip 0.2truecm The dependence of
the number of e-folds $\bar{\cal N}$ on the initial value $z_0$ of
the solutions of Eq.~(\ref{zg}).
    \hskip 1truecm}
    \label{Fig6_}
\end{figure}
The dependence of the number of e-folds on the initial value $z_0$
is shown in Fig.~\ref{Fig6_}. Here one can see that the
most of the solutions of Eq.~(\ref{kE0}) have big values of
$\bar{\cal N}$. One can reach $\bar{\cal N}\sim 50-100$ for $z_0
\sim 10-15$, i.e., without any fine tuning. An approximate form of
this dependence (at a few percent level of accuracy) for
$p_{{}_\Phi}=1/250$ is given by the following expression, obtained
by a numerical analysis of the solutions of Eq.~(\ref{kE0}): \ben
\bar{\cal N}(z_0) \approx 2.552\,z_0^{-3/10}+0.090\,z_0^{8/3}.
\la{NzApp} \een

\begin{figure}[htbp]
\vspace{5.5truecm} \includegraphics{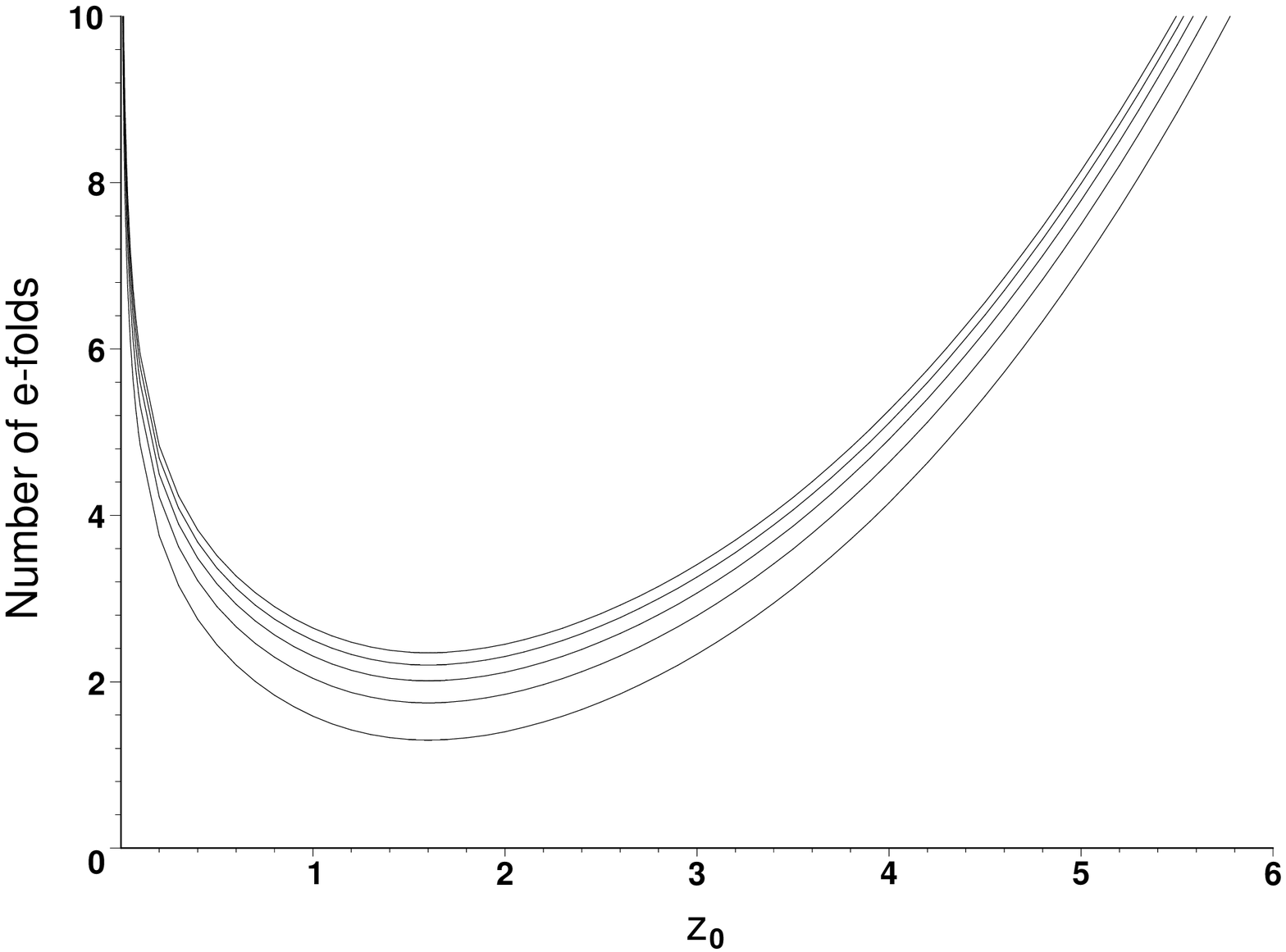}

\caption{\hskip 0.2truecm The dependence of the number of e-folds
$\bar{\cal N}$ on the initial value $z_0$ of the  solutions of
Eq.~(\ref{zg}). The curves for different values $p_{{}_\Phi}=1/50,
1/100, 1/150, 1/200, \,\hbox{and}\,\,\, 1/250$ are shown. As seen
in Fig.\ref{Fig6}, for $p_{{}_\Phi}\to 0$ the family of curves
converges to a limiting curve, which bounds this family from
above.
    \hskip 1truecm}
    \label{Fig6}
\end{figure}

Another important numerical observation is the dependence of the
relation $\bar{\cal N}(z_0,p_{{}_\Phi})$ on the value of the
parameter $p_{{}_\Phi}$. As seen in Fig.~\ref{Fig6}, there is a
good convergence of the corresponding family of curves, when
$p_{{}_\Phi}\to 0$. Having in mind that in the realistic case
$p_{{}_\Phi} \leq 10^{-30}$, we see that the inflation in 4D-DG is
a robust property with respect to variations of the mass of the
dilaton. The position of the minimum of the number of e-folds is at
$z_0\approx 1.6$ and seems to be independent on this mass. The
value of the minimum is $\gtrapprox 2.5$ in the limiting case
$p_{{}_\Phi}\to 0$.

This qualitative consideration, combined with the structure of the
phase portrait shown in Figures \ref{Fig1} and \ref{Fig2}, not
only explains the universal character of the inflation in 4D-DG,
but gives us a better understanding of its basic characteristics.
In particular, we see that we do not need fine tuning of the model
to describe the inflation as a typical physical phenomenon.

Using Eqs.\ (\ref{N}), (\ref{TN}), and (\ref{D00}), we obtain \ben
{\cal N}_{infl}^{(i)}(p_{{}_\Phi}^2)={\cal N}_{infl}^{(i)}(0)
+{\cal O}(p_{{}_\Phi}^2),
\hskip 0.2truecm \nonumber \\
t_{infl}^{(i)}(p_{{}_\Phi})= {\frac {4}
{\sqrt{3}}}p_{{}_\Phi}\Theta_{infl}^{(i)}(0) +{\cal
O}(p_{{}_\Phi}^3),\nonumber \een where \ben {\cal
N}_{infl}^{(i)}(0)=\int_{{\cal C}_{in}^{fin}(i)}\limits {\frac
{dg/g}{{\cal D}(g,z;0)}},\,\,\,\hbox{and}
\nonumber \\
\Theta_{infl}^{(i)}(0)=\int_{{\cal C}_{in}^{fin}(i)}\limits {\frac
{z^2 g\,dg}{ {\cal D}(g,z;0)} }, \hskip .15truecm \la{Gamma} \een
are independent of the parameter $p_{{}_\Phi}$. Taking into
account the extremely small physical value of this parameter, one
can conclude that the higher order terms in ${\cal
N}_{infl}^{(i)}(p_{{}_\Phi}^2)$ and $\Delta
t_{infl}^{(i)}(p_{{}_\Phi})$ are not essential. Neglecting them,
we actually ignore the contribution of the term $\Phi/3$ in the
functions $Z_{1,2}$ (\ref{Z1Z2}), or the corresponding term
$\Phi^2$, in the cosmological potential $U(\Phi)$
(\ref{Vsimplest}). It is natural to ignore these terms in the
domain of inflation, because they are essential only in a small
vicinity of dSV. In the function $W(\Phi)$, we have a term that
dominates for $\Phi-1 \gg p_{{}_\Phi}^2$, having a huge
coefficient $\sim p_{{}_\Phi}^{-2}$. Physically this approximation
means that we are neglecting the small pure cosmological constant
term in the cosmological potential and preserve only the terms,
which are proportional to the mass of the dilaton.

The relation
$\Delta t_{infl}^{(i)}(p_{{}_\Phi}^2)\sim
{\frac 4 {\sqrt{3}}}p_{{}_\Phi} \Theta_{infl}^{(i)}$,
written in physical units, reads
\ben
E_{{}_\Phi} \Delta t_{infl}^{(i)}\sim
\hbar {\frac 4 {\sqrt{3}}}\Theta_{\cal N}^{(i)}(0).
\la{mT}
\een
It resembles some kind of a quantum
``uncertainty relation'' for the rest energy
$E_{{}_\Phi}$ of the dilaton and the time of inflation
and probably indicates the quantum character
of the inflation as a physical phenomenon.

More important for us is the fact that Eq.~(\ref{mT})
shows the relationship between the mass of the dilaton,
$m_{{}_\Phi}$, and the time duration of the inflation.
Having large enough mass of the dilaton, we will have small
 time duration of the inflation.
This recovers the real meaning of the mass $m_{{}_\Phi}$
as a physical parameter in 4D-DG, and gives possibility
to determine it from astrophysical observations as
a basic cosmological parameter.

%%%%%%%%%%%%%%%%%%%%%%%%%%%%%

\section{Concluding Remarks}

In this section, we discuss some open problems in 4D-DG.

The main open physical problem at the moment seems to be the
precise determination of the dilaton mass. The known restriction
on it is too weak. It is convenient to have a dilaton $\Phi$ with
mass $m_{{}_\Phi}$ in the range $10^{-3}$--$10^{-1}\,{\rm eV}$. In
this case, the dilaton will not be able to decay into the other
particles of the Standard Model, since they would have greater
masses \cite{E-F_P}. On the other hand, in 4D-DG we do not need
such a suppressing mechanism since a direct interaction of the
dilaton $\Phi$ with matter of any kind is forbidden by the weak
equivalence principle \cite{F02}. This gives us the freedom to
enlarge significantly the mass of the dilaton without
contradiction with the known physical experiments. One of the
important conclusions of the present article is that $m_{{}_\Phi}$
is related to the time duration of the inflation. One is tempted
to try a new speculation -- to investigate a 4D-DG with dilaton
mass, $m_{{}_\Phi}$, in the domain $100\,{\rm GeV}$--$1\,{\rm
TeV}$, and dilaton Compton wave length, $l_{{}_\Phi}$, between
$10^{-18}$ and $10^{-16}\,{\rm cm}$. In this case the
time-duration of the inflation, $t_{{}_\Phi}$, will be of order of
$10^{-28}\,{\rm sec}$; the dimensionless dilaton parameter,
$p_{{}_\Phi}$, will be about $10^{-45}$, and the ultra-high
frequency, $\omega_{{}_\Phi}$, of dilatonic oscillations during de
Sitter asymptotic regime will be approximately $10^{19}\,{\rm
GHz}$. Such new values of the basic dilaton parameters are very
far from the Planck scales. They seem to be accessible for the
particle accelerators in the near future, and raise new physical
problems.

Another open problem is the justification of quantum corrections
and the exact form of cosmological potential. More general
potentials than (\ref{Vsimplest}) where introduced in \cite{F02},
and further justification and verification is needed. One must
take into account the possible additional corrections in order to
obtain a correct physical description in the early Universe.

Important open problems in the development of a general
theoretical framework of 4D-DG are the detailed theory of
cosmological perturbations, structure formation, and possible
consequences of our model for the CMB parameters. The properties
of the solutions of the basic equation for linear perturbations
$\delta\Phi$ differ essentially from the ones in other
cosmological models with one scalar field. This equation yields a
strong ``clusterization'' of the dilaton $\Phi$ at very small
distances \cite{E-F_P}. Actually, the equation for dilaton
perturbations shows the existence of ultra-high frequency
oscillations, described above, and non-stationary gravi-dilaton
waves with length $l_{{}_\Phi}\leq 10^{-2}\,{\rm cm}$. Such new
phenomena cannot be viewed as a clusterization at astrophysical
scales. Thus, their investigation as unusual cosmological
perturbations is an independent interesting issue. For example, it
is interesting to know whether it is possible to consider these
space-time oscillations as a kind of dark matter in the Universe.

A more profound description of the inflation in 4D-DG,
both in the absence and in the presence of matter and
space-curvature, is needed.
It requires correct averaging of dilatonic oscillations.

We intend to present the corresponding results elsewhere.

\begin{acknowledgments}

One of us (PF) is deeply grateful to Professor Steven Weinberg for
his invitation to join the Theory Group at the University of Texas
at Austin where an essential part of this work was done and for
his kind hospitality. This visit was supported by Fulbright
Educational Exchange Program, Grant Number 01-21-01.

We are grateful to Dr.\ Nikola Petrov and to Dr.\ Roumen Borissov
who carefully read the manuscript and made numerous remarks that
enhanced the clarity of the exposition.

This research was supported in part by NSF grant PHY-0071512, and
by the University of Sofia Research Fund, contract 404/2001.

\end{acknowledgments}

\end{document}